\documentclass[english,aps,twocolumn,showpacs,superscriptaddress]{revtex4-1}
\usepackage[utf8]{inputenc}
\usepackage{hyperref}
\usepackage{amsmath}
\usepackage[normalem]{ulem}
\usepackage{bm}
\usepackage{bbm}
\usepackage{color}
\usepackage{natbib}
\usepackage{graphicx}

\begin{document}

\begin{abstract}
We report sub-nanometer, high-bandwidth measurements of the out-of-plane (vertical) motion of atoms in freestanding graphene using scanning tunneling microscopy (STM). By tracking the vertical position over a long time period, a thousand-fold increase in the ability to measure space-time dynamics of atomically thin membranes is achieved over the current state-of-the-art imaging technologies. We observe that the vertical motion of a graphene membrane exhibits rare long-scale excursions characterized by both anomalous mean-squared displacements and Cauchy-Lorentz power law jump distributions.
\end{abstract}

\title{Anomalous dynamical behavior of freestanding graphene membranes
}
\author{M.L. Ackerman}
\affiliation{Department of Physics, University of Arkansas, Fayetteville, AR 72701, USA}
\author{P. Kumar}
\affiliation{Department of Physics, University of Arkansas, Fayetteville, AR 72701, USA}
\author{M. Neek-Amal}
\affiliation{Department of Physics, University of Antwerp, Groenenborgerlaan 171, B-2020 Antwerpen, Belgium}
\author{P.M. Thibado}
\email[]{thibado@uark.edu}
\affiliation{Department of Physics, University of Arkansas, Fayetteville, AR 72701, USA}
\author{F.M. Peeters}
\affiliation{Department of Physics, University of Antwerp, Groenenborgerlaan 171, B-2020 Antwerpen, Belgium}
\author{Surendra Singh}
\affiliation{Department of Physics, University of Arkansas, Fayetteville, AR 72701, USA}
\date{\today}

\maketitle
Stochastic processes are ubiquitous in nature. Their studies have played a pivotal role in the development of modern physics and provided the first evidence of the atomic nature of matter\cite{einstein1905}.  Langevin initiated a truly dynamical theory for Brownian motion by conceiving a stochastic differential equation of motion for the particle\cite{langevin1908}. This model, often called Ornstein-Uhlenbeck model, predicts mean squared displacement MSD$(\tau)\propto \tau$, a velocity autocorrelation function (VCAF) exponentially decaying in time, and a Maxwell-Boltzmann equilibrium velocity distribution\cite{li2010}. Recent advances in measurement precision and resolution have extended the framework of Brownian motion to unprecedented space-time scales and to a wider variety of systems, including atomic diffusion in optical lattices and spin diffusion in liquids\cite{stapf1995,katori1997}. Studies of such systems are providing insights into the mechanisms and interactions responsible for stochasticity. For example, the particle may execute classical Brownian motion in a small neighborhood, but then move suddenly over a large distance to a new neighborhood, where it resumes classical movement. This is the crux of L\'evy walks with finite speeds and finite waiting times, in which the higher-velocity segments and jump lengths of the movement yield long-tailed power law distributions\cite{shlesinger1993,metzler2000}.  It has been hypothesized that L\'evy walks are present in a diverse set of systems, ranging from economics, biomedical signals, climate dynamics, and even animal foraging. It is now believed that an optimized search algorithm, even within information foraging theory, should utilize a L\'evy stable distribution with infinite variance\cite{vishwanathan1999}.

Membrane fluctuations, characterized by movement perpendicular to the membrane's surface, also fall under the purview of Brownian motion. Bio-membranes, in which thermal fluctuations aid the transport of chemicals through channels to the interior of a cell, \cite{kosztin2004} have been studied experimentally using nuclear magnetic resonance spectroscopy and optical microscopy\cite{bocian1978,pecreaux2004}. Moreover, modern theories of membrane structure and dynamics, which include elasticity as well as stochastic effects via Langevin equation, predict a Maxwell-Boltzmann distribution for the local fluctuations of the membrane \cite{naji2009,reister2010}. At present there are no direct experimental observations to test these predictions. 

Freestanding graphene is an ideal crystalline membrane that can be probed without degradation on an atomic scale with STM in an ultrahigh vacuum (UHV) environment. Using this approach, it was shown that the ripples can be described using an Ising model by mapping curved up (down) ripples into up (down) states of an Ising spin \cite{schoelz2015}.  An essential component missing from these studies is a measurement of the dynamic fluctuations. Here, we use STM to track the movement of a single carbon atom-sized region of the fluctuating membrane with sub-nanometer resolution.  We show that the membrane executes Brownian motion with rare large height excursions indicative of L\'evy walks. In addition, the membrane velocity obeys a long-tail Cauchy-Lorentz power law distribution, rather than a Maxwell-Boltzmann distribution. 

Monolayer graphene, commercially grown on Ni (less than 10 percent is multi-layer graphene), was directly transferred to a 2000-mesh, ultrafine copper grid having a lattice of square holes 7.5 microns wide with bar supports 5 microns wide. Scanning electron microscope (SEM) images show 90 percent coverage\cite{xu2012}. An Omicron ultrahigh-vacuum (base pressure is $10^{-10}$ mbar) low-temperature  STM, operated at room temperature, was used for the height measurements. The graphene film was mounted toward the sample plate on stand-offs, so the STM tip approached through the holes of the grid, in order to provide a more stable support. The entire STM chamber rests on an active, noise cancelling, vibration isolation system and is powered using a massive battery bank with an isolated building ground to achieve exceptionally low electrical noise.

Data was acquired using STM tips fabricated in-house, under constant-current (feedback on) tunneling conditions, and the topography scan set to point-mode (no x- or y-scanning). We adapted our system to allow us to continuously record 16-bit data for both the actual tunneling current and the tip height at a rate of 800 Hz for a time span of $10^4$ seconds, yielding 8 million data points per channel. We independently monitored the STM tip-sample drift, and found that it is non-stochastic and less than 1 nm/hr. We collected data from multiple membranes for fixed imaging conditions spanning several orders-of-magnitude in tunneling current (0.01 nA to 10 nA) and bias voltage ( 0.01 V to 10 V), all at room temperature. When imaging the graphene surface with atomic resolution, we observe only monolayer graphene that is free of defects over a scale of microns.

\begin{figure}[htbp]
\begin{center}
\includegraphics[width=3.5in]{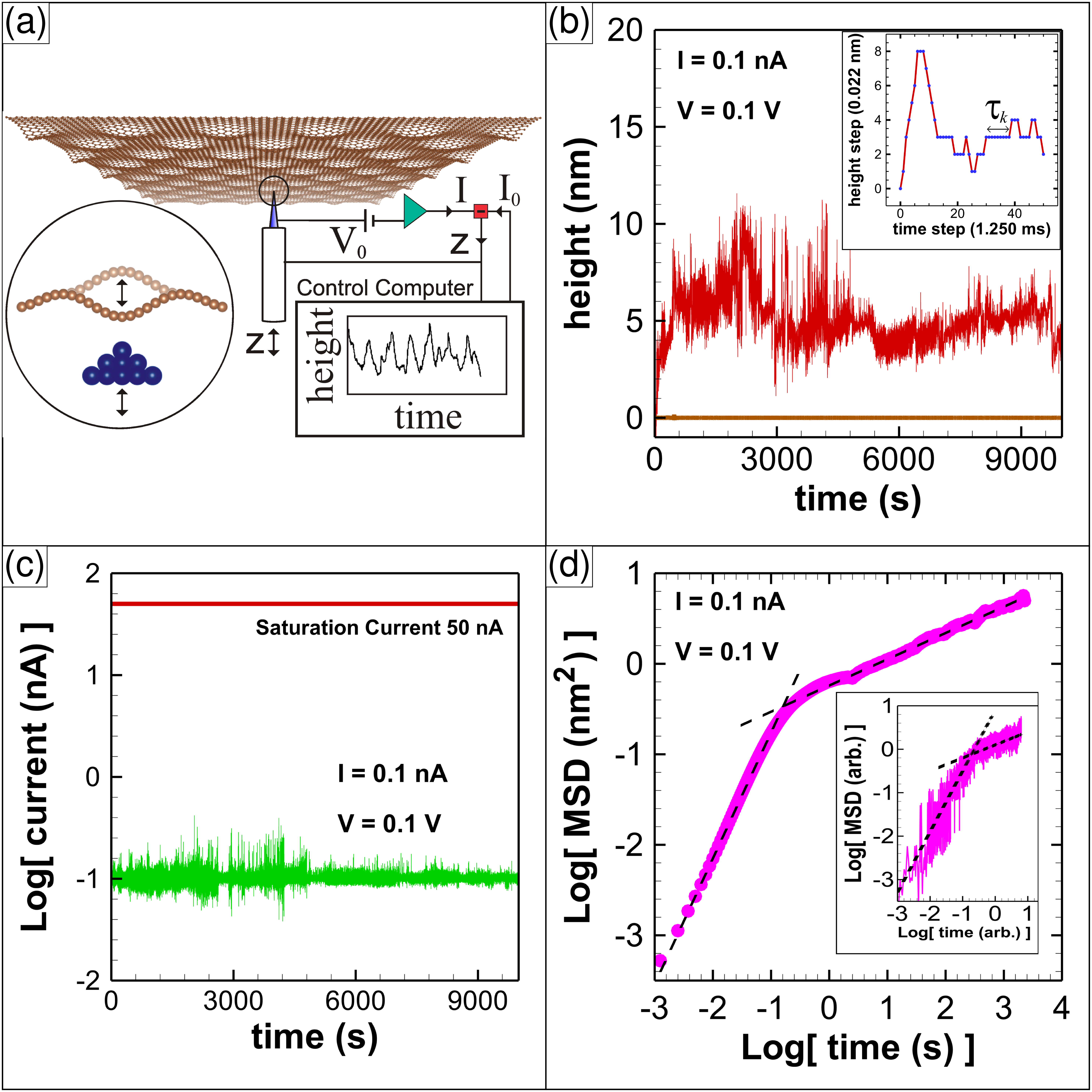}
\caption{\small (a) Outline of experimental setup. (b) Typical time trace of membrane height (above) and from a rigid sample (below). The inset is an expanded view of the freestanding graphene time trace.  (c) Typical tunneling current profile during the measurement. (d) Mean squared displacement (MSD) of membrane height as a function of time. Dashed lines are fits with slopes 1.4 and 0.3. The inset is the result of a simulation using exponential wait times and Cauchy jump lengths. Again, the dashed lines are fits with slopes 1.4 and 0.3.}
\label{fig1}
\end{center}
\end{figure}

A simplified schematic of our experimental setup is shown in Fig. 1(a). A biased STM tip, mounted at the end of a piezoelectric tube scanner, approaches the electrically grounded freestanding graphene membrane from below. A typical time series for $z(t)$ is shown in Fig. 1(b) for STM set point ($I$=0.1 nA, $V$=0.1 V). The range of membrane movement ($\sim$ 10 nm) is enormous for point-mode STM, and for comparison a typical STM trace acquired from a rigid sample is also shown. Such large values of $z(t)$ appear reasonable as the unsupported graphene membrane forms a rippled structure that shifts continuously between a large number of energetically equivalent configurations \cite{meyer2007, los2009}. The inset in Fig. 1(b) shows a zoomed-in plot of membrane height (in units of  $\delta_o =0.022$ nm) as a function of time (in units of $\tau_o =1.250$ ms) with the typical time between two successive jumps (i.e., a change in the height) labeled as $\tau_k$. The wait-time probability distribution for this data was calculated and follows a simple exponential, showing that it is a Poisson process. The measured tunneling current in Fig. 1(c), corresponding to the data shown in Fig. 1(b), remained well below the saturation level and well above zero, even when the membrane height changed significantly. The contribution of tip-sample distance variation to the membrane height $z(t)$ was negligible for all of our data. In addition, the cross-correlation coefficient between the measured height and tunneling current is less than 0.05.

From the time series $z(t)$, we computed its mean-squared displacement MSD$(\tau)\equiv \langle \big(z(t+\tau)-z(t)\big)^2\rangle$, which is shown in Fig. 1(d).  This data, spanning nearly 7 orders-of-magnitude in time, is characterized by a power law dependence of MSD$(\tau)\sim \tau^\alpha$ with $\alpha\ne1$ being the anomalous diffusion exponent. For our data, the motion at short times is characterized by $\alpha=1.4$ (super-diffusive motion) followed by a range for which $\alpha=0.3$ (sub-diffusive motion). We observe the same exponents with other data sets acquired at different tunneling setpoints. A random walk simulation, using exponential wait times and Cauchy jump lengths, yields a MSD with super-diffusion for short times and sub-diffusion for long times, as shown in Fig. 1(d) inset. Experimental evidence for a Cauchy distribution also comes from the membrane velocity. 

\begin{figure}[htbp]
\begin{center}
\includegraphics[width=3.5in]{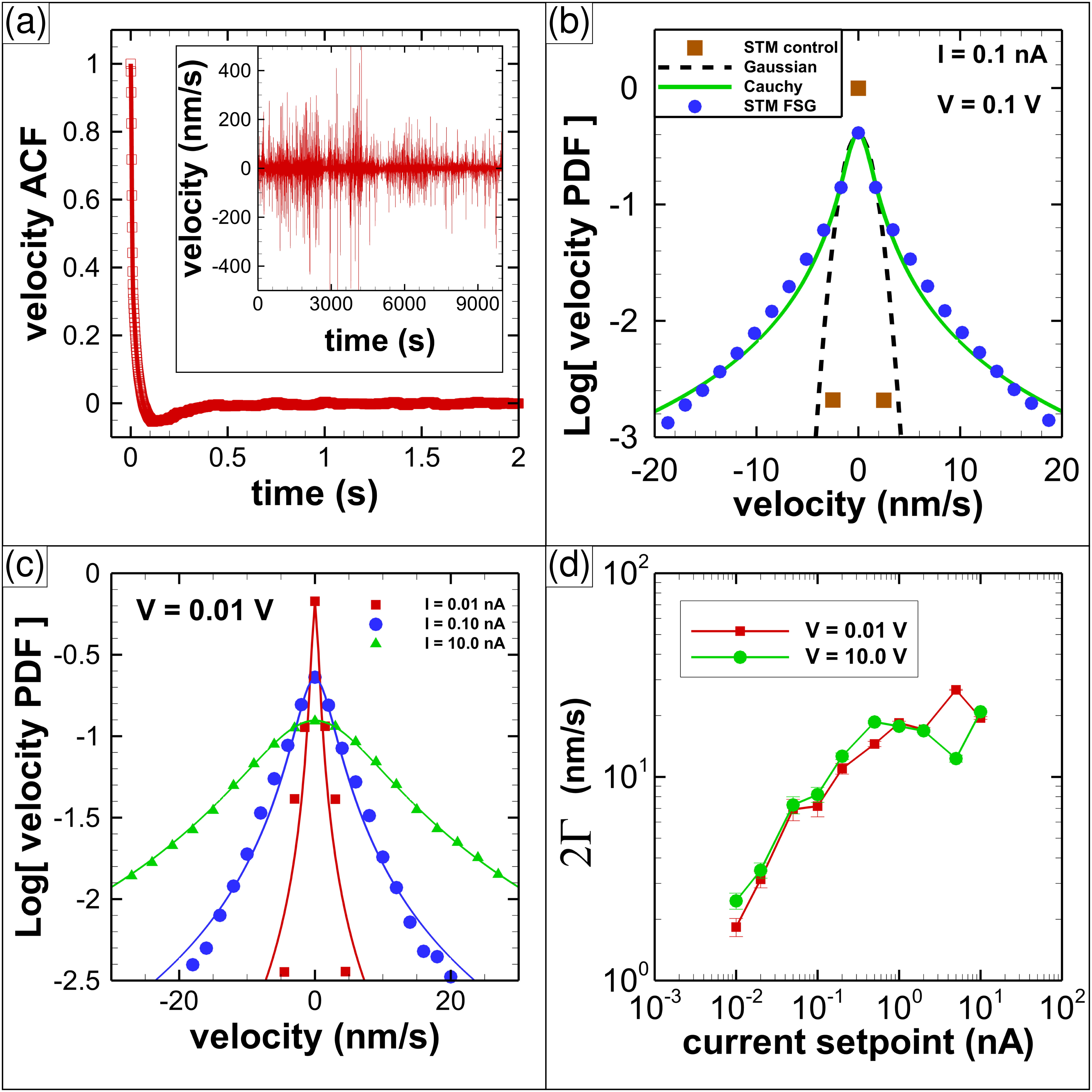}
\caption{\small (a) Velocity autocorrelation function and instantaneous velocity (inset) computed from membrane height $z(t)$ shown in Fig. 1(b). (b) Measured freestanding graphene (FSG) membrane velocity probability distribution function (PDF) fitted to Cauchy-Lorentz and Gaussian distributions, along with the rigid control sample (square symbols). (c) Velocity PDFs and Cauchy-Lorentz fits (full curves) for different tunneling currents. (d) Variation in the FWHM of the velocity PDFs with tunneling current for two different bias voltage setpoints.}
\label{fig2}
\end{center}
\end{figure}

Instantaneous membrane velocity, computed numerically from the time series for $z(t)$, shown in Fig. 1(b), displays highly irregular behavior [inset of Fig. 2(a)] with a short memory.  The velocity autocorrelation function VACF$(\tau)\equiv\langle v(t)v(t+\tau)\rangle$ for our data is shown in Fig. 2(a). It decreases rapidly, becoming negative around 0.1 s, indicative of a liquid-like behavior, before finally decaying to zero (within 0.5 s of our 10,000 s long measurement), showing that the membrane velocity fluctuations  are quickly decorrelated. This observation clearly demonstrates that it is possible to measure the equilibrium velocity distribution using STM, which derives support from other studies of single-atom diffusion using STM\cite{swartzentruber1996}.

Figure 2(b) shows the membrane velocity probability distribution function (PDF) computed from the data shown in Fig. 1(b).
The solid curve is the best-fit Cauchy-Lorentz distribution with zero mean velocity, $v_o=0$ and full width at half maximum (FWHM) $2\Gamma$
\begin{equation}
{\cal L}(v;\Gamma)=\frac{1/\pi\Gamma}{1+ [(v-v_o)/\Gamma]^2}\,.
\end{equation}
The velocity distribution peaks at zero and is symmetric about it, consistent with an equal likelihood of the membrane moving up or down (i.e., balanced movement in the presence of the STM tip), and indicates that we are within the elastic limit for our bias voltage setpoints.

Even though membrane velocities as high as 500 nm/s were observed, 98\% of all velocities fall in the range ($-15,\,15$) nm/s. The dashed curve is the best fit Gaussian. The data clearly follow a Cauchy-Lorentz distribution rather than a Gaussian, especially for speeds greater than 10 nm/s. Note, the three square data points are for the rigid sample data shown in Fig. 1(b), and are shown for comparison. Figure 2(c) shows the membrane velocity PDF data for tunneling currents spread over many orders of magnitude along with best fit Cauchy-Lorentz distributions. These velocity distributions were obtained from STM data taken from a new location on the sample with increasing tunneling current. The most striking conclusion is that in all cases, the membrane velocities follow a Cauchy-Lorentz distribution \cite{comment1} (i.e., a L\`evy stable distribution with infinite variance and stability index 1) much better than a Gaussian. 

It can further be seen from Fig. 2(c), that the velocity distribution broadens with increasing tunneling current.  The same trend is apparent in Fig. 2(d), which shows the variation of the FWHM of the velocity PDF with the tunneling current set-point over the entire range of this study. The broadening of the distribution is consistent with Joule heating due to the STM tunneling current providing more kinetic energy to the membrane \cite{comment2}.

Elasticity theory predicts the vibrational modes of freestanding graphene, however, it does not yield information about the stochastic processes. Atomistic simulations can provide insight into the observed phenomenon. For our molecular dynamics (MD) simulations, we prepared a pre-buckled, curved down square membrane (15 nm by 15 nm) containing 10,000 carbon atoms with boundary atoms fixed (no STM tip). The simulations in vacuum were performed in LAMMPS using the AIREBO potential~\cite{lammps}. Nos\'e-Hoover thermostat was used to maintain a constant temperature and the equations of motions were integrated using a time-step of $1$~fs. System was first equilibrated for $0.5$~ns starting from the initial configuration and the subsequent trajectory from a production run of $1$~ns was used for the analysis.

\begin{figure}[htbp]
\begin{center}
\includegraphics[width=3.5in]{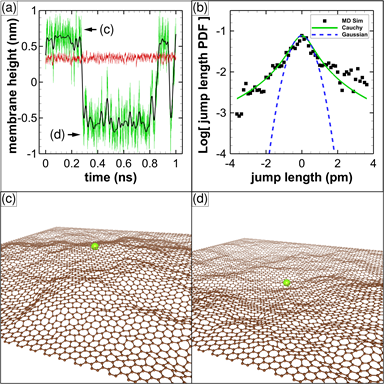}
\caption{\small (a) Height of the central carbon atom in time from MD simulation for low temperatures (100K) and high temperatures (3000K). The high temperature data is found to transition from positive to negative heights four times over $1$~ns. A low-pass filtered height is also shown. (b) The jump length probability distribution function (PDF) for the low-pass filtered height data is shown with a best fit to Cauchy-Lorentz and Gaussian distributions. (c) Perspective view of the membrane in a curved down shape marked as ``(c)" in Fig. 3(a). (d) Perspective view of the membrane in a curved up marked as ``(d)" in Fig. 3(a).}
\label{fig3}
\end{center}
\end{figure}

The movement of the central atom with $10^6$ time steps (1 per fs) at low-temperature (100 K) shows ~0.1 nm height fluctuations at an overall height of ~0.35 nm above the fixed boundary atoms as shown in Fig. 3(a). At higher temperatures (3000 K) something significantly different happens. At the same time scale, the random movement results in mirror buckling of the entire membrane from above the fixed boundary atoms to below them. Fig. 3(c) and (d)  shows two snapshots of the membrane for opposite configurations labeled (c) and (d) on Fig. 3(a). The long excursion from curved-down to curved-up is indicative of L\`evy walks. In fact,  if we average the data in Fig. 3(a) over a short time interval to smooth out rapid fluctuations (black curve in Fig. 3(a), we always obtain a Cauchy jump length distribution, as shown in Fig. 3(b). Time-averaging of the very high frequency movement of graphene is exactly what our STM measurement would yield. Similar Cauchy distributions are also obtained if we spatial average about the central atom, which would also naturally occur with any real measurement having resolution greater than 1 nm.

This large-scale movement is a consequence of graphene changing locally its overall curvature (e.g., a curved down to curved up transition), while the small-scale movements are simple vibrations of the membrane with no inversion of its curvature. We can track how this happens in the high temperature simulation, the random up and down movement at times add together in the same direction resulting in a long-excursion to another equilibrium configuration on the other side of the fixed boundary atoms. Given that four such events happen in 1 ns at 3000K, one can predict that these events will happen several times for our STM measurements carried out at room temperature\cite{voter2000}. Previously, we reported that the presence of a temperature gradient can induce mirror buckling\cite{neek-amal2014}. In this manuscript, we report a new mechanism, spontaneous mirror buckling, which occurs without a temperature gradient.

Our measurements uncover an unexplored spatial and temporal domain in membrane fluctuations with profound implications both for our fundamental understanding and technological applications of membranes. Properly understood, the random membrane fluctuations can be usefully exploited. For example, energy harvesting from the continuous movement of a massive system is an important application of stochastic nano-resonators\cite{gammaitoni}.

By tuning the velocity distribution (as we demonstrated by varying the tunneling current), one can activate certain processes and deactivate others. Furthermore, as the membrane flexes, it modifies the local strain, the chemical reactivity, and the charge distribution, which allows the system to do work. Engineering specific channel geometries in a membrane, along with complementary ratchet-style components, could create small artificial L\'evy motors\cite{haenggi2009}. Finally, advances in our understanding of membrane dynamics will help us to control the motion of objects over the membrane, which is critical to protein function, as well as the self-organization of artificial materials.  

In summary, dynamics of atomic scale fluctuations of a freestanding graphene membrane were studied using point-mode scanning tunneling microscopy, and molecular dynamics simulations. Our measurements reveal the richness of the random out-of-plane motion of membranes, which exhibit anomalous dynamics and long-tail equilibrium distributions of dynamical variables symptomatic of L\'evy walks. We also demonstrated  that the stochastic properties of fluctuating membranes can be controlled using STM.  This, coupled with the ability to observe motion with atomic-scale resolution, provides an ideal system to study new Brownian motion regimes and test various models of anomalous transport. In conclusion, we experimentally and theoretically demonstrate that buckling events in 2D materials yield artificial crystalline membranes with tunable Levy walks. It is important to emphasize that without the two breakthroughs presented in this study, we would be unable to reach these conclusions \cite{xu2014}. Ultimately, this study provides methods to predict, control, and, even minimize the occurrence of large scale, sudden changes in a wide variety of systems. 

\bibliographystyle{apsrev4-1}

\begin{thebibliography}{99}
\bibitem{comment1}It might be that a truncated Cauchy-Lorentz distribution is a better fit but that will not change the conclusion here.
\end{thebibliography}

\textbf{Acknowledgements}

The authors thank Theodore L. Einstein, Michael F. Shlesinger, and Woodrow L. Shew for their careful reading of the manuscript and insightful comments. This work was supported by the Flemish Science Foundation (FWO-Vl) and the Methusalem Foundation of the Flemish Government. P.M.T. was supported by the Office of Naval Research under Grant No. N00014-10-1-0181 and the National Science Foundation under Grant No. DMR-0855358.

\end{document}